\documentclass[aps,prb,twocolumn,groupedaddress,showpacs]{revtex4}

\usepackage{graphicx}
\usepackage{tabularx}
\usepackage{amssymb}
\usepackage{amsmath}
\usepackage{bm}
\usepackage{subfigure}
\usepackage{graphpap}
\usepackage{multirow}
\usepackage{notes2bib}
\usepackage[version=3]{mhchem} 

\begin{document}
\title{Bound excitons and many-body effects in x-ray absorption spectra of azobenzene-functionalized self-assembled monolayers}
\author{Caterina \surname{Cocchi}}
\affiliation{Institut f\"ur Physik and IRIS Adlershof, Humboldt-Universit\"at zu Berlin, Berlin, Germany}
\affiliation{European Theoretical Spectroscopic Facility (ETSF)}
\email{caterina.cocchi@physik.hu-berlin.de}
\author{Claudia \surname{Draxl}}
\affiliation{Institut f\"ur Physik and IRIS Adlershof, Humboldt-Universit\"at zu Berlin, Berlin, Germany}
\affiliation{European Theoretical Spectroscopic Facility (ETSF)}
\date{\today}
\pacs{71.35.-y, 73.20.Mf, 78.70.Dm}
\begin{abstract}
We study x-ray absorption spectra of azobenzene-functionalized self-assembled monolayers (SAMs), investigating excitations from the nitrogen $K$ edge.
Azobenzene with H-termination and functionalized with \ce{CF3} groups is considered.
The Bethe-Salpeter equation is employed to compute the spectra, including excitonic effects, and to determine the character of the near-edge resonances.
Our results indicate that core-edge excitations are intense and strongly bound: Their binding energies range from about 6 to 4 eV, going from isolated molecules to densely-packed SAMs.
Electron-hole correlation rules these excitations, while the exchange interaction plays a negligible role.
\end{abstract}
\maketitle
\section{Introduction}
Azobenzene-functionalized self-assembled monolayers (SAMs) on metal surfaces represent a viable and efficient way to obtain ordered architectures of photo-switching molecules. \cite{jasc+96jpc,wang+97jec,evan+98lang,mann+02jpcb,akiy+03jpcb,schm+08apa,ludw+13pccp}
However, it has been observed that in such closely packed systems photo-isomerization can be drastically hindered by steric effects, \cite{tama+02lang,kuma+08nl,weid+08lang,frey+09jppac,jung+10lang,vall+13lang} and even by excitonic coupling between the chromophores. \cite{gahl+10jacs} 
In order to overcome these limitations and obtain SAMs with efficient switching rates, a number of strategies have been developed, such as modifying the morphology of the substrate, \cite{mold+15lang} introducing organic spacers, \cite{jung+12jpcc,jaco+14pccp} and functionalizing azobenzene with end groups.\cite{bret+12jpcm}

\begin{figure}
\centering
\includegraphics[width=.45\textwidth]{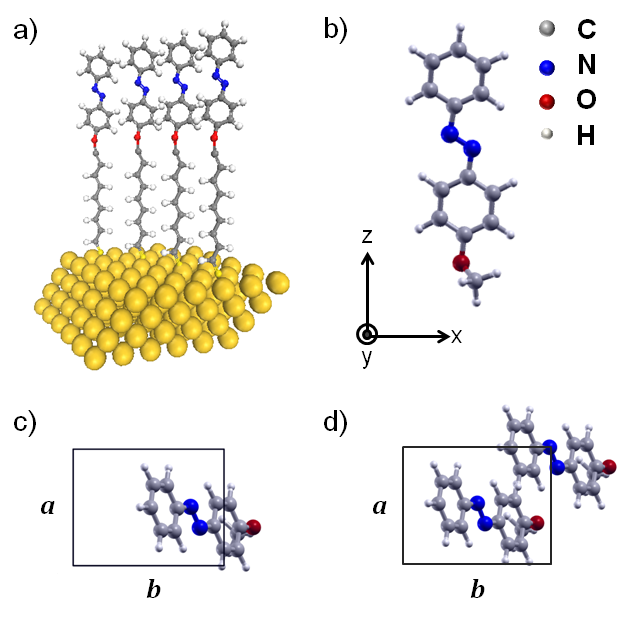}%
\caption{(Color online) a) Sketch of azobenzene-functionalized SAM of alkanethiols on a gold surface, b) isolated azobenzene molecule, c) diluted (d-SAM) and d) packed SAM (p-SAM) in their unit cells, including one and two inequivalent molecules, respectively.
}
\label{figure1_H-az}
\end{figure}

To tune these complex systems in view of optimized performance, a deep knowledge of their chemical composition and structure-property relationship  is required.
x-ray absorption spectroscopy (XAS) represents a powerful technique for this purpose, and a synergistic interplay with theory can provide an insightful interpretation of the experimental data.
First-principles methods represent the most suitable tool.
Density-functional theory (DFT), both in the \textit{core-hole} approximation \cite{tana-mizo09jpcm}  and the $\Delta$-self-consistent-field ($\Delta$SCF) approach, \cite{besl+09jcp} is routinely applied to simulate XAS in a wide range of materials, from gas-phase molecules to solid-state systems. \cite{nybe+99prb,urqu-ade02jpcb,kolc-herm04ss,lask+09jpcm,hua+10prb}
Recently, also time-dependent DFT has become popular to compute core-level excitations in molecular compounds. \cite{tu+07pra,besl-asmu10pccp,schm+10jcp,nard+11pccp,lopa+12jctc}
While these approaches can provide qualitative agreement with experiments, explicit many-body treatment has turned out superior to such approaches.
In small molecules, coupled-cluster methods have been successfully applied to compute XAS from the carbon and nitrogen K-edge. \cite{cori+12pra,fran+13jcp}
For solid-state materials, many-body perturbation theory (MBPT) represents the state-of-the-art formalism to describe neutral excitations.\cite{shir00jesrp,shir04jesrp}
The electron-hole (\textit{e-h}) interaction, effectively described by the Bethe-Salpeter equation (BSE), plays a crucial role not only in conventional semiconductors, \cite{rohl-loui00prb,bott+07rpp} but also in organic crystals \cite{ruin+02prl,pusc-ambr02prl,humm+04prl,ambr+09njp} and even in isolated molecules. \cite{cocc-drax15condmat}
A number of studies on core-level excitations, from different edges and in several materials, \cite{olov+09prb,olov+09jpcm,lask-blah10prb,olov+11prb,vins+11prb,vins-rehr12prb,olov+13jpcm,nogu+15jctc} has demonstrated that BSE can accurately reproduce XAS.

In this paper, we present an \textit{ab initio} study of x-ray absorption spectra of azobenzene-functionalized SAMs.
We consider excitations from the nitrogen (N) $K$ edge, i.e., involving transitions from 1$s$ electrons to the conduction bands.
In this manner, we obtain exciton binding energies and determine the character of the core-level excitations.
Going from the isolated molecule to a closely-packed SAM, we analyze the XAS at increasing density of azobenzene molecules, and we compare our results with experimental data.
In order to understand whether and how functional groups affect the nature of the excitons and their binding energy, we consider molecules that are either  H-terminated (H-az) or functionalized with trifluoromethyl (\ce{CF3}-az).

%
\section{System and Methods}
\subsection{Azobenzene SAMs}
A sketch of an azobenzene-functionalized SAM of alkanethiols on gold is presented in Fig. \ref{figure1_H-az}a.
The chromophores are covalently bonded to the alkyl chains, which are attached to the gold substrate through a thiol group.
As suggested by scanning tunneling microscopy (STM) measurements, \cite{wolf+95jpc,jasc+96jpc} the SAM has an orthorhombic supercell, with lattice vectors $a$=6.05 \AA{} and $b$=7.80 \AA{}, hosting two inequivalent azobenzene molecules.
In our calculations, we neglect the alkyl chains and the gold surface, since they are expected not to play a role in the XAS from the N $K$ edge.
Therefore, we consider only the azobenzene molecule, with a methoxy group added to one end (see Fig. \ref{figure1_H-az}b), in order to reproduce the chemical environment of the covalent bond to the alkyl chain.
The reciprocal distance and orientation of the molecules in the unit cell is set according to the STM data. \cite{wolf+95jpc,jasc+96jpc}
Although the first experiments on these systems predicted a herringbone structure of the chromophores in the SAMs, \cite{wang+96cl} a consensus about the orientation of azobenzenes is still missing.
We consider the two inequivalent molecules in the unit cell being oriented parallel to each other, since we expect deep core levels to be hardly affected by the reciprocal orientation of the molecules.
In this configuration (see Fig. \ref{figure1_H-az}d) the azobenzenes are separated by about 2 \AA{} in the lateral direction, and by $\sim$ 3.8 \AA{} in the direction perpendicular to the plane of the phenyl rings.
We incorporate $\sim$ 14 \AA{} of vacuum in the vertical direction, to effectively simulate a two-dimensional system.

In order to understand the effects of packing in the XAS, we consider an additional structure, including only one molecule in the same unit cell.
We refer to this system, shown in Fig. \ref{figure1_H-az}c, as \textit{diluted} SAM (d-SAM), to distinguish it from the \textit{packed} SAM (p-SAM, Fig. \ref{figure1_H-az}d).
For comparison, we investigate an isolated azobenzene molecule in an orthorhombic supercell, with $\sim$ 6 \AA{} of vacuum in each direction.
We also consider SAMs of \ce{CF3}-functionalized azobenzene, which have been recently synthesized. \cite{gahl+10jacs}
\ce{CF3} and other functional groups are used in experiments as markers, to identify the orientation of the molecules with respect to the surface, \cite{gahl+10jacs} and/or to tune the switching properties of the SAMs by decreasing the steric hindrance due to intermolecular interactions. \cite{bret+12jpcm} 
Also for \ce{CF3}-az, we investigate p- and d-SAMs, as well as an isolated molecule for comparison. 
We adopt the same structures shown for H-az in Fig. \ref{figure1_H-az}.

\subsection{Theoretical Background}
\label{section:theory}
x-ray absorption spectra are computed from first principles by solving the BSE, which is an effective equation of motion for the electron-hole two-particle Green's function. \cite{hank-sham80prb,stri88rnc}
By considering only transitions from core ($c$) to unoccupied ($u$) states, the BSE in matrix form reads:
\begin{equation}
\sum_{c'u'\mathbf{k'}} \hat{H}^{BSE}_{cu\mathbf{k},c'u'\mathbf{k'}} A^{\lambda}_{c'u'\mathbf{k'}} = E^{\lambda} A^{\lambda}_{cu\mathbf{k}} .
\label{eq:BSE}
\end{equation}
In case of N $K$ edge, the N 1$s$ is the only initial state.
The BSE Hamiltonian in Eq. \ref{eq:BSE} can be written as the sum of three terms: 
\begin{equation}
\hat{H}^{BSE} = \hat{H}^{diag} + 2 \gamma_x \hat{H}^x + \gamma_c \hat{H}^{dir}.
\label{eq:H_BSE}
\end{equation}
The \textit{diagonal} term $\hat{H}^{diag}$ accounts for single-particle transitions.
Including only this term corresponds to the independent-particle approximation (IPA).
The \textit{exchange} ($\hat{H}^x$) and \textit{direct} ($\hat{H}^{dir}$) terms in Eq. \ref{eq:H_BSE} incorporate the repulsive bare and the attractive screened Coulomb interaction, respectively.
The coefficients $\gamma_x$ and $\gamma_c$ in Eq. \ref{eq:H_BSE} enable to select the \textit{spin-singlet} ($\gamma_x$ = $\gamma_c$ = 1) and \textit{spin-triplet} ($\gamma_x$ = 0, $\gamma_c$ = 1) channels.
In the latter case, the exchange interaction is not present.
In Eq. \ref{eq:BSE}, the eigenvalues $E^{\lambda}$ represent excitation energies.
Exciton binding energies ($E_b$) are defined, for each excitation, as the difference between excitation energies $E^{\lambda}$ computed from IPA and BSE.
The eigenvectors $A^{\lambda}$ carry information about the character and composition of the excitons.
Through the transition coefficients
\begin{equation}
\mathbf{t}_{\lambda}= \sum_{cu\mathbf{k}} A^{\lambda}_{cu\mathbf{k}} \dfrac{\langle c\mathbf{k}|\widehat{\mathbf{p}}|u\mathbf{k}\rangle}{\epsilon_{u\mathbf{k}} - \epsilon_{c\mathbf{k}}} ,
\label{eq:t}
\end{equation}
$A^{\lambda}$ enter the expression of the imaginary part of the macroscopic dielectric function ($\varepsilon_M$):
\begin{equation}
\mathrm{Im}\varepsilon_M = \dfrac{8\pi^2}{\Omega} \sum_{\lambda} |\mathbf{t}_{\lambda}|^2 \delta(\omega - E_{\lambda}) .
\label{eq:ImeM}
\end{equation}
%

\subsection{Computational Details}
\label{section:comput}
All calculations are performed with the \texttt{exciting} code, \cite{gula+14jpcm} a computer package implementing DFT and MBPT. \cite{sagm-ambr09pccp}
\texttt{exciting} is based on the all-electron full-potential augmented planewave method, which ensures an explicit and accurate description of core electrons.
The calculation of XAS via the solution of the BSE in an all-electron framework has been successfully applied to different absorption edges in a number of bulk materials, \cite{olov+09prb,olov+09jpcm,lask-blah10prb,olov+11prb,olov+13jpcm} including, very recently, small molecules. \cite{nogu+15jctc}

The Kohn-Sham (KS) electronic structure, used here as starting point for the BSE, is computed within the local-density approximation (Perdew-Wang functional). \cite{perd-wang92prb}
Quasiparticle energies are approximated by KS single-particle energies, and a scissors operator is applied to match the experimental absorption onset for the p-SAM, according to the available data for H-az \cite{mold+15lang} and \ce{CF3}-az. \cite{gahl+10jacs}
The same correction is applied also to the respective isolated molecule and d-SAM, since we do not have experimental data available for these systems. \bibnote{Larger self-energy corrections can be expected for the molecules. However, we are studying trends and comparing binding energies, i.e., energy differences. Hence, our assumption is fully justified.}
A $\mathbf{k}$-point mesh of 6$\times$4$\times$1 (3$\times$2$\times$1) is used to sample the Brilloiun zone for the p-SAM (d-SAM), in both ground-state and BSE calculations. 
For the basis functions, a planewave cutoff $\mathbf{G}_{max}$=5 bohr is applied to the molecules; for the SAMs it is reduced to 4.625 bohr.
Muffin-tin spheres of radii $R_{MT}$=0.8 bohr are considered for hydrogen, $R_{MT}$=1.1 bohr for nitrogen and fluorine, and $R_{MT}$=1.2 bohr for carbon and oxygen.
The atomic positions of each structure are optimized by minimizing the Hellmann-Feynman forces within a threshold of 0.025 eV/\AA{}.
For the calculation of the $e$-$h$ interaction term $H^{dir}$ in the BSE (Eq. \ref{eq:H_BSE}), the screening is evaluated within the random-phase approximation, including the N 1$s$ core states, all valence states and 100 conduction bands.
Local-field effects are included, with at least 400 $|\mathbf{G}+\mathbf{q}|$ vectors for the SAMs and about 2000 for the molecules.
These parameters ensure convergence of the XAS within 0.25 eV.

\section{Results and Discussion}

\begin{table*}
\begin{tabular}{c||c c|c c|c c||c|c|c}
\textit{Method} & \multicolumn{6}{c||}{BSE} & \multicolumn{3}{c}{IPA} \\ \hline \hline
\textit{Peak} & \multicolumn{2}{c|}{A'} & \multicolumn{2}{c}{B'} & \multicolumn{2}{|c||}{C'} & A & B & C \\ \hline \hline
Molecule & 396.3 & [\textbf{6.0}] & 400.8 & [\textbf{3.7}] & 401.2 & [\textbf{4.2}] & 402.3 & 404.5 & 405.4 \\ \hline
d-SAM & 396.8 & [\textbf{5.7}] & 401.3 & [\textbf{3.3}] & 401.7 & [\textbf{3.6}] & 402.5 & 404.6 -- 404.8 & 405.3 -- 405.6 \\ \hline
p-SAM & 398.0 & [\textbf{4.1}] & 402.0 & [\textbf{1.9}] & 402.4 & [\textbf{2.9}] & 402.1 -- 402.9 & 403.9 -- 405.0 & 405.3 -- 405.9 \\ \hline \hline
 \end{tabular} 
\caption{Excitation energies of the main peaks of the spectra presented in Fig. \ref{figure2_H-az_XAS}, for molecule and SAMs. A scissors operator of 26.8 eV is applied to the underlying KS electronic structures to match the experimental absorption onset (Ref. \onlinecite{mold+15lang}). Exciton binding energies, corresponding to the difference between IPA and BSE excitation energies (i.e., $E_A - E_{A'}$, $E_B - E_{B'}$, and $E_C - E_{C'}$), are highlighted in bold. All quantities are expressed in eV.}
\label{table1}
\end{table*}

\begin{figure}
\centering
\includegraphics[width=.45\textwidth]{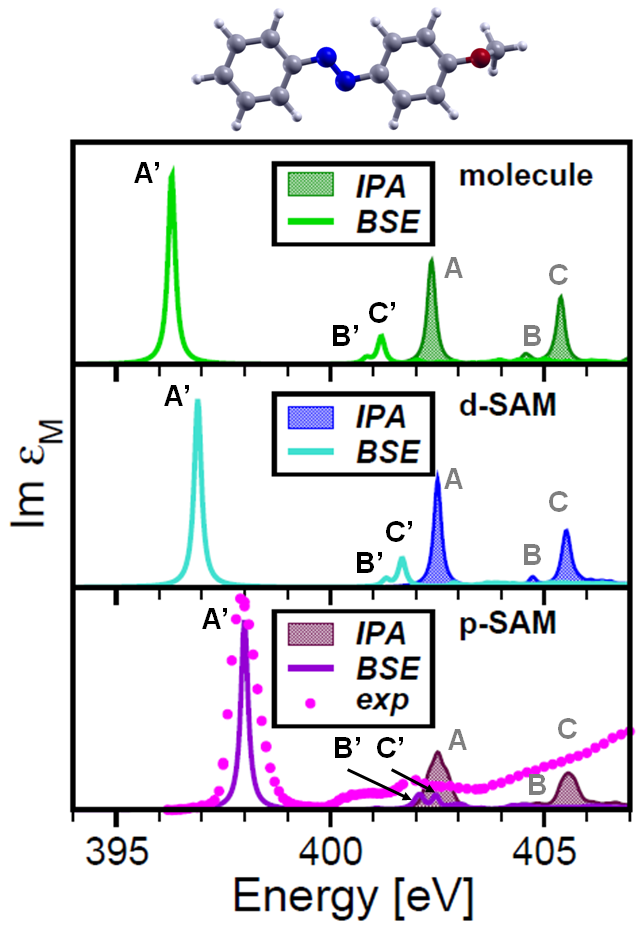}%
\caption{(Color online) XAS of isolated azobenzene molecule (\textit{top}), d-SAM (\textit{middle}), and p-SAM (\textit{bottom}). In the BSE spectra (solid line), the peaks are labelled as A', B' and C'. Independent-particle spectra (IPA -- solid area) are shown for comparison (peaks A, B and C). The imaginary part of the macroscopic dielectric function is averaged over the three Cartesian components. The experimental curve for the p-SAM is taken from Ref. \onlinecite{mold+15lang}. A Lorentzian broadening of 0.1 eV is applied to the theoretical spectra. H-az is shown on top.
}
\label{figure2_H-az_XAS}
\end{figure}

\begin{figure}
\centering
\includegraphics[width=.4\textwidth]{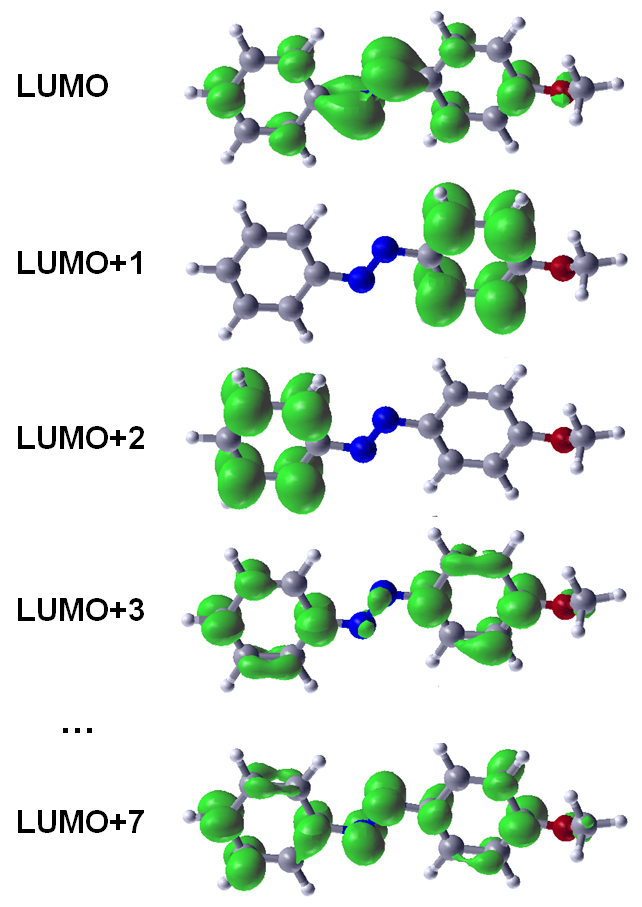}%
\caption{(Color online) Lowest unoccupied molecular orbitals of the azobenzene molecule. An isovalue of 0.002 is used for the isolsurfaces.
}
\label{figure3_figure3_H-az_MOs}
\end{figure}

In Fig. \ref{figure2_H-az_XAS}, we present the XAS computed for the azobenzene molecule and for the p- and d-SAMs.
In addition to the result obtained from the solution of the BSE, the corresponding IPA spectrum is shown in each panel.
To better guide the reader, we start our analysis from the IPA results.
These are presented only for comparison with the BSE spectra, to better highlight excitonic effects.
Three peaks, labeled as A, B and C, can be identified approximately at the same energy in each structure, independently of the packing density.
Since each molecule has two N atoms, with the 1$s$ levels separated by 61 meV in the KS spectrum, this multiplicity is reflected also in the XAS.
Hence, in the spectrum of the isolated molecule (\textit{top} panel) each peak is formed by two transitions.
The first and most intense peak, A, corresponds to a transition to the LUMO level, which presents $\pi^{*}$ character, with strong localization on the azo group (see Fig. \ref{figure3_figure3_H-az_MOs}).
Also the other peaks, B and C, are given by $\pi^{*}$ resonances, involving excitations to LUMO+3 and LUMO+7, respectively.
By inspecting the orbitals, shown in Fig. \ref{figure3_figure3_H-az_MOs}, we observe a direct correspondence between the amount of charge localized on the azo group and the strength of the resonance.
Peaks A and C have large intensity, while B is rather weak.
Between A and B additional transitions to the unoccupied levels LUMO+1 and LUMO+2 take place.
Considering that these orbitals are localized predominantly on the phenyl rings (Fig. \ref{figure3_figure3_H-az_MOs}), the corresponding transitions are extremely weak.

The independent-particle picture for the SAMs presents analogous features observed for the isolated molecule. 
Interestingly, not only the relative intensity of the three peaks is the same, but also their energy is independent of the packing density.
In the SAMs a larger number of transitions contribute to the peaks, compared to the single molecule. 
This is especially true for the p-SAM, where the peaks A and C experience a broadening of about 1 eV and 0.5 eV, respectively (see Table \ref{table1}).
In the spectrum of the d-SAM, 8 single-particle transitions contribute to the lowest-energy peak A, which is comprised within less than 0.1 eV.
Also the nature of the transitions is the same in the single molecule and in the SAMs.

The scenario changes significantly when $e$-$h$ interaction is taken into account.
By inspecting the BSE spectrum of the single molecule (Fig. \ref{figure2_H-az_XAS}, \textit{top} panel), we observe again three peaks, labeled as A', B' and C', whose energies are significantly red-shifted compared to their IPA counterparts (see Table \ref{table1}).
The exciton corresponding to the peak A' has a sizable binding energy of 6.0 eV.
Also B' and C' are remarkably red-shifted compared to their IPA counterparts, by 3.7 eV and 4.2 eV, respectively. 
Interestingly, the $e$-$h$ interaction is stronger in C' than in B'.
It is worth noting that the relative energy difference between A' and B' (4.6 eV) is more than twice than the one between A and B (2.2 eV).
Also the energy difference between A' and C' (4.9 eV) exceeds by about 50$\%$ the one between A and C (3.1 eV).
In contrast, B' and C' are separated only by 0.4 eV, which is half of the difference between B and C (0.9 eV).
Also the oscillator strengths are considerably redistributed upon inclusion of the $e$-$h$ interaction.
A' is almost twice more intense than A, with a consequent decrease of relative spectral weight of B' and C'.
From the analysis of the exciton composition, we observe that A' corresponds to a pure transition to the LUMO, as within the IPA.
Conversely, B' presents a rather mixed character, with a dominant contribution ($\sim$70$\%$) from the transition to the LUMO+3 state.
Also C' is composed by transitions to LUMO+7 ($\sim$ 70$\%$) and LUMO ($\sim$12$\%$).

The main features observed in the XAS of the single molecule can be recognized also for the SAMs (Fig. \ref{figure2_H-az_XAS}, \textit{middle} and \textit{bottom} panels).
The lowest-energy peak A' dominates the spectrum in terms of spectral weight, and its energy is red-shifted with respect to A by 5.7 and 4.1 eV in the d- and p-SAM, respectively (see also Table \ref{table1}).
In the case of the d-SAM, the peaks B' and C' are significantly red-shifted with respect to their IPA counterparts B and C, by 3.3 and 3.6 eV, respectively.
Like for the molecule, also in this case the binding energy of C' is larger than the one of B'.
In the spectrum of the p-SAM, $E_b \simeq$ 3 eV for C', and $E_b \lesssim$ 2 eV for B' (see Table \ref{table1}).
These values of binding energies, especially for the core-edge exciton A', are remarkably large \cite{Note5} with respect to optical excitations in molecular crystals, where typically $E_b \lesssim$ 1 eV. \cite{humm-ambr05prb}
They are also large compared to inorganic materials: For example, near-edge resonances from the Li or Be $K$ edge in binary crystals present binding energies ranging from a few hundreds of meV up to $\sim$ 2 eV. \cite{olov+09jpcm,olov+13jpcm}

The BSE spectra in Fig. \ref{figure2_H-az_XAS} are almost rigidly blue-shifted of about 2 eV, when going from the isolated molecule to the p-SAM. 
Considering that the IPA absorption onset, given by the position of A, is approximately the same in the three systems (see also Table \ref{table1}), the decreasing binding energy of A' upon increasing azobenzene packing density is driven by many-body interactions.
We assign this effect to a combination of enhanced screening and wave-function delocalization, as observed in optical excitations of organic crystals.\cite{ruin+02prl,humm+03prb,pusc+03prb}
Moreover, as a consequence of dipole-dipole coupling, the plethora of single-particle transitions forming the lowest-energy peak in the SAMs, combine into two excitons, corresponding to the peak A'. \bibnote{In the case of the p-SAM, having two inequivalent molecules in the unit cell, each exciton has multiplicity 2.}

The predominant role of the $e$-$h$ correlation in such core-edge excitons is further confirmed by an additional analysis of the BSE results.
When computing triplet excitation energies, we diagonalize a BSE Hamiltonian with $\gamma_x =0$ (Eq. \ref{eq:H_BSE}), since triplet states do not experience exchange interaction.
By comparing singlet and triplet excitation energies, we notice differences of the order of 20 -- 50 meV for all the considered excitons.
This result is independent of the packing density of the molecules and reveals that the screened Coulomb interaction is the driving mechanism of the XAS.
Conversely, local-field effects (LFE), ruled by the exchange term $H^x$, are irrelevant here, due to the localized character of the excitation. 
There is in fact a very small overlap between the $h$, localized in the N 1$s$ state, and the $e$ in the conduction region.
This represents a rather different scenario than optical absorption, where LFE are significantly larger. 
In particular, organic materials are typically characterized by singlet-triplet splitting of the same order of magnitude of singlet binding energies. \cite{rohl-loui99prl,pusc-ambr02prl,humm-ambr05prb,ambr+06cp}
Concerning the character of the excitons in the SAMs, the picture is not significantly different from that of isolated azobenzene.
The exciton A' corresponds to transitions from N 1$s$ to the lowest unoccupied band.
In the p-SAM, which has two inequivalent molecules in the unit cell, this band is split in two. 
Transitions to both subbands mix up to form the exciton.
The peak B' is given by a manifold of weak excitations, with a rather mixed character, in both d- and p-SAM.
They include transitions to unoccupied states, which are the counterparts of LUMO+3 in the single molecule (see Fig. \ref{figure3_figure3_H-az_MOs}).
The peak C' is composed by only 2 and 4 excitons in the d- and p-SAM, respectively. 
These excitons have a remarkably mixed character, and involve transitions to higher unoccupied bands.

\begin{figure}
\centering
\includegraphics[width=.45\textwidth]{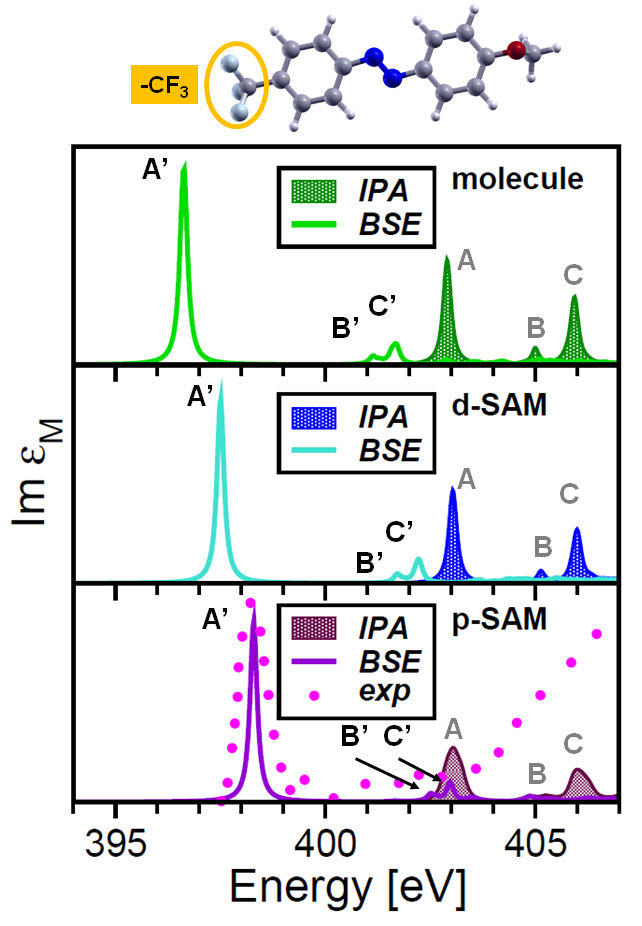}%
\caption{(Color online) XAS of isolated \ce{CF3}-functionalized azobenzene molecule (\textit{top}), d-SAM (\textit{middle}), and p-SAM (\textit{bottom}). In the BSE spectra (solid line), the peaks are labelled as A', B' and C'. Independent-particle spectra (IPA -- solid area) are shown for comparison (peaks A, B and C). The imaginary part of the macroscopic dielectric function is averaged over the three Cartesian components. The experimental curve for the p-SAM is taken from Ref. \onlinecite{gahl+10jacs}. A Lorentzian broadening of 0.1 eV is applied to the theoretical spectra. \ce{CF3}-az is shown on top.
}
\label{figure4_CF3-az_XAS}
\end{figure}

The quality of our BSE results is confirmed by the comparison with available experimental data for the p-SAM. \cite{mold+15lang}$^,$ \bibnote{Our theoretical results are compared with angle-resolved measurements. For a quantitative comparison, we refer here to data obtained for the so-called \textit{magic angle}, for which the spectrum is independent of the light polarization. Intensities are normalized to the height of the core-edge peak.}
By inspecting Fig. \ref{figure2_H-az_XAS}, we notice that the experimental spectrum is characterized by a strong peak, corresponding to A'.
It is attributed to a $\pi^*$ resonance, due to the 1$s$ $\rightarrow$ LUMO transition: \cite{mold+15lang}  this is in perfect agreement with our finding.
About 4 eV above, a shoulder appears in correspondence of B' and C'.
Due to the limited resolution of the experimental spectrum, it is not possible to identify these two excitons separately.
However, also in the experiment, they are assigned to transitions to LUMO+$n$ states, having again $\pi^*$ character, \cite{mold+15lang} as confirmed by our results.
It is worth noting that besides the peaks A', B' and C', the experimental spectrum has non-zero intensity around 401 eV.
This feature is present also in our BSE spectrum with extremely small oscillator strength.\cite{Note6}
This weak peak is given by two double degerate excitations, targeting the two lowest unoccupied bands (LUMO and LUMO+1).
Furthermore, in the high-energy part of the experimental spectrum, the oscillator strength increases towards $\sim$ 408 eV, where a bump due to the 1$s$ $\rightarrow$ $\sigma^*$ transition is observed. \cite{mold+15lang}
This feature is not reproduced by theory, since $\sigma^*$ states, which lie very high in energy in the KS spectrum, are not considered in the solution of the BSE.

\begin{table*}
\begin{tabular}{c||c c|c c|c c||c|c|c}
\textit{Method} & \multicolumn{6}{c||}{BSE} & \multicolumn{3}{c}{IPA} \\ \hline \hline
\textit{Peak} & \multicolumn{2}{c|}{A'} & \multicolumn{2}{c}{B'} & \multicolumn{2}{|c||}{C'} & A & B & C \\ \hline \hline
Molecule & 396.6 & [\textbf{6.3}] & 401.1 & [\textbf{3.9}] & 401.6 & [\textbf{4.4}] & 402.9 & 405.0 & 406.0 \\ \hline
d-SAM & 397.5 & [\textbf{5.5}] & 401.7 & [\textbf{3.5}] & 402.2 & [\textbf{3.7}] & 403.0 & 405.2 & 405.9 -- 406.1 \\ \hline
p-SAM & 398.3 & [\textbf{4.5}] & 402.5 & [\textbf{2.1}] & 403.0 & [\textbf{2.8}] & 402.8 -- 403.4 & 404.6 -- 405.7 & 405.8 -- 406.4 \\ \hline \hline
 \end{tabular} 
\caption{Excitation energies of the main peaks of the spectra presented in Fig. \ref{figure4_CF3-az_XAS}, for \ce{CF3}-az molecule and SAMs. A scissors operator of 27.4 eV is applied to the underlying KS electronic structures to match the experimental absorption onset (Ref. \onlinecite{gahl+10jacs}). Exciton binding energies, corresponding to the difference between IPA and BSE excitation energies (i.e., $E_A - E_{A'}$, $E_B - E_{B'}$, and $E_C - E_{C'}$), are highlighted in bold. All quantities are expressed in eV.}
\label{table2}
\end{table*}

Finally, we investigate the x-ray absorption spectra of azobenzene molecules and SAMs terminated with a trifluoro-methyl group (\ce{CF3}-az, see Fig. \ref{figure4_CF3-az_XAS}).
Since no additional N atoms are introduced in the system, we expect to observe the same features as in the XAS of the H-terminated counterparts.
With this analysis we aim at understanding the influence of functionalization on binding energies and exciton character.
We again consider an isolated \ce{CF3}-functionalized azobenzene molecule, as well as d- and p-SAMs, in order to inspect the role of packing density.
The calculated XAS are shown in Fig. \ref{figure4_CF3-az_XAS}, and the (binding) energies of the bright excitons are summarized in Table \ref{table2}.
The spectra appear strikingly similar to those presented in Fig. \ref{figure2_H-az_XAS} for H-terminated azobenzene, and so the main features analyzed previously.
The XAS are considerably blue-shifted from the molecule to the p-SAM, i.e., upon increasing intermolecular interactions. 
The intense peaks correspond to transitions to unoccupied states, having the same $\pi^*$ character as in the H-az systems. 
By inspecting carefully Tables \ref{table1} and \ref{table2}, we notice that exciton binding energies slightly increase in presence of \ce{CF3} termination.
This functional group has an electron-withdrawing character and introduces a sizable dipole moment in the molecule, of the order of 5 Debye.
This slightly enhances the $e$-$h$ attraction, thus strengthening the exciton binding energy of the main peaks in Fig. \ref{figure4_CF3-az_XAS}, of about 0.2 eV on average.
In the p-SAM the binding energy of A' is 0.4 eV larger than in the H-az system. 
On the contrary, in the d-SAM the value of $E_b$ decreases by 0.2 eV for the lowest-energy resonance A', compared to its H-terminated counterpart.
In a similar fashion, the optical absorption onset computed for polycyclic aromatic hydrocarbons is red-shifted by about 0.3--0.5 eV, in presence of edge-functional groups.\cite{cocc+12jpcc}
Also for \ce{CF3}-az, we observe singlet-triplet splittings ($\sim$50 meV), which are two orders of magnitude smaller than the binding energies.
The comparison with the experimental data from Ref. \onlinecite{gahl+10jacs} indicates good agreement with our results.
In both theoretical and experimental spectra the intense low-energy peak A', as well as the weaker resonances B' and C', present the same energy separation and relative intensity.
Since $\sigma^*$ states are not included in our BSE calculation, the bump above 405 eV is not reproduced in our spectrum.

\section{Summary and Conclusions}
We have investigated N 1$s$ x-ray absorption spectra of azobenzene-functionalized SAMs, determining the nature of the excitations and discussing the role of many-body effects.
Our results, obtained from \textit{ab initio} calculations, in the framework of many-body perturbation theory, reveal a clear excitonic character of the main peaks in the XAS.
Binding energies for core-edge excitons, computed from the solution of the Bethe-Salpeter equation, decrease from 6 eV in the molecule to 4 eV in packed SAMs. 
This is a many-body effect assigned to an interplay between screening and exciton delocalization.
Based on this finding, we expect exciton coupling between different molecules to be even more pronounced in the optical range, where $e$-$h$ pairs are typically more delocalized. 
This could give insight into the loss of switching capability, as observed for densely packed SAMs. \cite{gahl+10jacs}

Core-level excitations in these systems are ruled by the attractive $e$-$h$ correlation.
The exchange interaction plays a negligible role, as testified by singlet-triplet splittings, which are two orders of magnitude smaller than exciton binding energies.
Functionalization with a \ce{CF3} group does not affect the overall spectral features, but only induces a slight increase in the exciton binding energies of $\sim$0.2 eV on average.
Good agreement is observed with available experimental data.

In conclusion, our work confirms the predictive power of many-body perturbation theory in determining the character of the resonances and in disclosing the microscopic mechanisms ruling the core excitation process.
This confirms the indispensable role of theory not only in interpreting the experimental data, but also in gaining further insight into the physics of core-level spectroscopy.

\section*{Acknowledgement}
This work was funded by the German Research Foundation (DFG), through the Collaborative Research Center SFB-658. C.C.~acknowledges support from the \textit{Berliner Chancengleichheitsprogramm} (BCP).

\end{document}